\documentclass[,final            
]
  {aipproc}

\layoutstyle{8x11single}

\begin{document}

\title{CMB data constraint on self-annihilation of dark matter particles}


\author{Jaiseung Kim}{address={Niels Bohr Institute, Blegdamsvej 17, DK-2100 Copenhagen, Denmark}}

\author{Pavel Naselsky}{address={Niels Bohr Institute, Blegdamsvej 17, DK-2100 Copenhagen, Denmark}}

\begin{abstract}
Recently, self-annihilation of dark matter particles is proposed to explain the ``WMAP Haze'' and excess of energetic positrons and electrons in ATIC and PAMELA results.
If self-annihilation of dark matter occurs around the recombination of cosmic plasma, energy release from self-annihilation of dark matter delays the recombination, and hence affects CMB anisotropy. By using the recent CMB data, we have investigated the self-annihilation of dark matter particles.
In this investigation, we do not found statistically significant evidence, and impose an upper bound on $\langle \sigma v \rangle/m_{\chi}$.
The upcoming data from Planck surveyor and the Fermi Gamma-ray telescope will allow us to 
break some of parameter degeneracy and improve constraints on self-annihilation of dark matter particles.
\end{abstract}

\maketitle


\section{Introduction}
According to the WMAP concordance model \citep{WMAP3:parameter,WMAP5:parameter}, cold dark matter constitutes $\sim 20\%$ of the total mass of our Universe.
There have been various studies on dark matter particles \cite{galactic_neutrino,heavy_neutrionos:cosmic_ray,heavy_neutrino,new_neutrino}.
However, there are still very little known about dark matter particles, since they interact with other ingredients (e.g. radiation, baryon) only through gravitation. In the WMAP data, the excess of emission is observed in the region around the Galactic center and is dubbed ``WMAP Haze''. Since ``WMAP Haze'' is uncorrelated with known Galactic foregrounds, it has been suggested that ``WMAP Haze'' might be synchrotron emission from energetic electrons and positrons produced in self-annihilation of dark matter particles  \citep{WMAP_Haze,WMAP_Haze_DM1,WMAP_Haze_DM2}. 
Self-annihilation of dark matter particles has been also suggested to explain the excess of the high energy positron and electron observed in PAMELA and ATIC data \citep{PAMELA,ATIC}.
If dark matter particles annihilate, extra energy would be injected into cosmic plasma.
Since the rest-mass energy of a dark matter particle is expected to be much higher than the ionization energy of hydrogen atoms, self-annihilation could affect the ionization history significantly. 
For the past years, there have been great successes in measurement of Cosmic Microwave Background (CMB) anisotropy by ground and satellite observations \citep{WMAP5:basic_result,WMAP5:powerspectra,WMAP5:parameter,ACBAR,ACBAR2008,QUaD1,QUaD2,QUaD:instrument}.
The five year data of the Wilkinson Microwave Anisotropy Probe (WMAP) \citep{WMAP5:basic_result,WMAP5:powerspectra,WMAP5:parameter} is released, and the recent ground-based CMB observations such as the ACBAR \citep{ACBAR,ACBAR2008} and QUaD \citep{QUaD1,QUaD2,QUaD:instrument} provide information complementary to the WMAP data. Recently, Planck surveyor \citep{Planck_bluebook} is successfully launched and going to measure CMB temperature and polarization anisotropy with very fine angular resolution.
Using the recent and future CMB data, we may impose significant constraints on cosmological models \citep{Modern_Cosmology,Inflation,Foundations_Cosmology}.
In particular, we may constrain the self-annihilation property of dark matter particles by investigating CMB data, since the CMB anisotropy is sensitive to the ionization history \citep{Delayed_recombination,Constraint_Recombination,Constraint_Recombination2,Ionization_history,Baryonic_recombination}.

In this paper, we investigate the effect of self-annihilating dark matter on the CMB power spectrum and constrain the self-annihilation of dark matter particles.
With respect to the previous studies \citep{CMB_WIMP,dm_tau,dark_decay,dm_Zhang,dm_Galli}, the numerical computer code, which is added to incorporate the self-annihilation effect, is slightly improved in accuracy. Besides that, our analysis includes more recent data (WMAP5 + QUaD + ACBAR), which include CMB temperature data of high angular resolution  and low noise polarization data.

The plan of this paper is as follows.
We discuss how self-annihilation of dark matter particles affect ionization history and CMB anisotropy.
Then, we constrain the self-annihilation of dark matter with the current CMB data, and present our analysis result.
Finally, we summarize our investigation and briefly discuss the prospect with the future observation.

\section{self-annihilation of dark matter and ionization history}
\label{ionization}
The evolution of free electron fraction $x_e$ over redshift $z$ satisfies:
\begin{eqnarray}
\frac{dx_e}{dz} =\frac{1}{(1+z)H(z)}[R_s(z)-I_s(z)-I_{X}(z)], \label{xe_evolution}
\end{eqnarray}
where $H(z)$ is Hubble rate.
On the right hand side of Eq. \ref{xe_evolution},
$R_s(z)$ and  $I_s(z)$ denote the standard recombination and ionization rate respectively.
The rate of ionization due to particle annihilation, $I_{X}(z)$, is given by \citep{energy_fraction_Shull,dark_decay,dm_Zhang,dm_Galli}:
\begin{eqnarray}
I_{X}(z)&=&C(1-x_e/3) \frac{dE/dt}{n_H(z)\,E_i}+(1-C)(1-x_e/3) \frac{dE/dt}{n_H(z) E_{\alpha}},\label{Ix}
\end{eqnarray}
where $E_i$ is the binding energy of a ground state hydrogen atom, 
$E_\alpha$ is the energy difference between 2p and 2s state of a hydrogen atom, and $n_H(z)$ is the number density of hydrogen nuclei. 
The factor $C$ in Eq. \ref{Ix} is given by \citep{Recombination_Peebles}:
\begin{eqnarray*}
C=\frac{1+K \Lambda_{2s1s}\,n_H(1-x_e)}{1+ K \Lambda_{2s1s}\,n_H(1-x_e) + K\,\beta_{B}\,n_H(1-x_e)}, 
\end{eqnarray*}
where $\Lambda_{2s1s}$ is the decay rate of 2s energy level into 1s level, $\beta_{B}$ is  photoionization rate, and 
$K=\lambda^3_{\alpha}/8\pi H(z)$ with the wavelength of Ly-$\alpha$ photon $\lambda_{\alpha}$.
Energy release rate due to dark matter annihilation is given by:
\begin{eqnarray}
\frac{dE}{dt}=5.6\;\rho^2_0 c^2 \Omega^2_{c}(1+z)^6\,F_{dm} 
\end{eqnarray}
where $\rho_0$ is the critical density of the present Universe, $\Omega_{c}$ is the density fraction of dark matter, and
\begin{equation}
F_{dm}=2 f \left(\frac{\langle \sigma v \rangle}{10^{-26}\mathrm{cm}^3 \mathrm{s}^{-1}}\right) \left(\frac{m_{\chi} c^2}{\mathrm{GeV}} \right)^{-1}. \label{Fdm}
\end{equation}
A fudge factor $f$ denotes the energy fraction deposited on baryonic gas ($0<f\le 1$), $m_{\chi}$ is the mass of a dark matter particle, and
$\langle \sigma v \rangle$ is the effective self-annihilation rate, where $\langle \sigma v \rangle$ is $\sim 10^{-26}\mathrm{cm}^3 \mathrm{s}^{-1}$ 
in the case of thermal relic abundance. 

\begin{figure}
\centering
\includegraphics[height=.3\textheight]{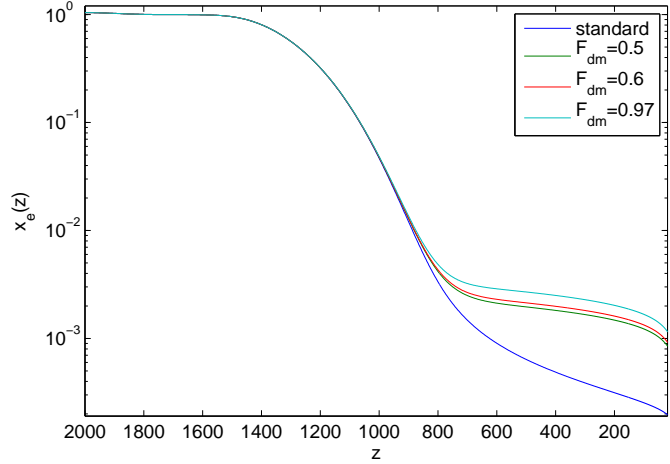}
\caption{Ionization history: fraction of free electrons, $x_e$, is plotted over a range of redshift $z$.}
\label{xe}
\end{figure}
By making a small modification to the widely used \texttt{RECFAST} code \cite{RECFAST1,RECFAST2,RECFAST3}, we have numerically computed the ionization history for various values of $F_{dm}$,  which are shown in Fig. \ref{xe}. 
We may see that self-annihilation of dark matter particles (i.e. $F_{dm}>0$) delays recombination.

\begin{figure}
\centering
\includegraphics[height=.3\textheight]{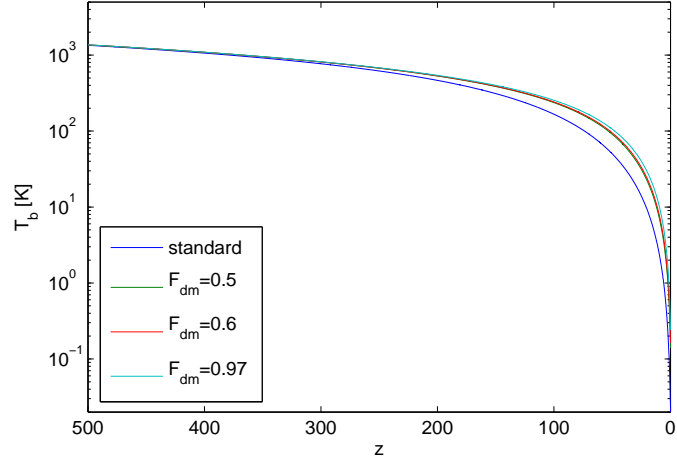}
\caption{Baryonic gas temperature: The temperature of baryonic gas, $T_b$, is plotted over a range of redshift $z$.}\label{Tb}
\end{figure}
Energy realease from self-annihilation of dark matter particles also affects the temperature of baryonic gas.
In the presence of self-annihilating dark matter, the evolution of baryonic gas temperature, $T_b$, is governed by \cite{dm_Galli,dark_decay,dm_Zhang}:
\begin{eqnarray}
(1+z)\frac{dT_b}{dz} =\frac{8 \sigma_T a_R T^4_{CMB}}{3 m_e cH(z)} \frac{x_e}{1+f_{He}+x_e}(T_b-T_{CMB}) +2 T_b-\frac{2}{3k_B H(z)} \frac{1+2x_e}{3}\frac{dE/dt}{n_H(z)(1+f_{He}+x_e)},\label{Tb_evolution}
\end{eqnarray}
where $\sigma_T$ is the Thompson scattering cross section, $f_{He}$ is the Helium fraction, $T_{CMB}$ is the photon temperature and $k_B$ is the Boltzmann constant.
The last term on the right hand side is associated with self-annihilation of dark matter.
Using a modified \texttt{RECFAST} code, we have numerically computed the evolution of baryonic gas temperature.
Fig. \ref{Tb} shows that self-annihilation of dark matter, as expected, increases baryonic gas temperature.

\section{CMB anisotropy}
\label{CMB}

Whole-sky Stokes parameters of CMB anisotropy are conveniently decomposed in terms of spin $0$ and spin $\pm2$ spherical harmonics: 
\begin{eqnarray*}
\Delta T(\hat{\mathbf n})&=&\sum_{lm} a_{T,lm}\,Y_{lm}(\hat{\mathbf n}),\\
Q(\hat {\mathbf n})\pm i U(\hat {\mathbf n})&=&\sum_{l,m} -(a_{E,lm}\pm i \,a_{B,lm})\;{}_{\pm2}Y_{lm}(\hat {\mathbf n}), 
\end{eqnarray*}
where $a_{T,lm}$, $a_{E,lm}$ and $a_{B,lm}$ are decomposition coefficients. 
In the standard inflation models, the decomposition coefficients satisfy the following statistical properties \cite{Modern_Cosmology,Inflation,Foundations_Cosmology,Cosmology}:
\begin{eqnarray}
\langle a_{T,lm}a^*_{T,l'm'}\rangle = C^{TT}_{l} \delta_{ll'}\delta_{mm'},\\
\langle a_{E,lm}a^*_{E,l'm'}\rangle = C^{EE}_{l} \delta_{ll'}\delta_{mm'},\\
\langle a_{T,lm}a^*_{E,l'm'}\rangle = C^{TE}_{l} \delta_{ll'}\delta_{mm'},
\end{eqnarray}
where $\langle \ldots \rangle$ denotes the average over an ensemble of universes.
In the absence of tensor perturbation, CMB power spectra and TE correlation are given by:
\begin{eqnarray}
C^{TT}_{l}&=&\frac{2}{\pi} \int k^2 dk\,P_{\Phi}(k)\:g^2_{Tl}(k), \label{ClTT}\\
C^{EE}_{l}&=&\frac{2}{\pi} \int k^2 dk\,P_{\Phi}(k)\:g^2_{El}(k),\label{ClEE}\\
C^{TE}_{l}&=&\frac{2}{\pi} \int k^2 dk\,P_{\Phi}(k)\:g_{Tl}(k)g_{El}(k),\label{ClTE}
\end{eqnarray}
where $g_{Tl}(k)$, $g_{El}(k)$ and $g_{Bl}(k)$ are the radiation transfer functions for corresponding modes, and $P_{\Phi}(k)$ denotes the power spectrum of primordial perturbation.
In most of inflationary models, primordial power spectrum $P_{\Phi}(k)$ nearly follows a power-law \citep{Inflation}, which makes good agreements with recent observations \citep{WMAP3:parameter,WMAP5:Cosmology}.

\begin{figure}
\includegraphics[height=.3\textheight]{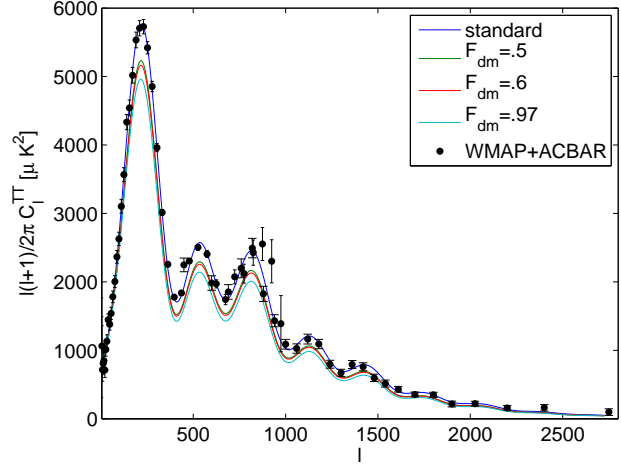}
\caption{CMB temperature power spectrum}
\label{Cl_TT}
\end{figure}

\begin{figure}
\includegraphics[height=.3\textheight]{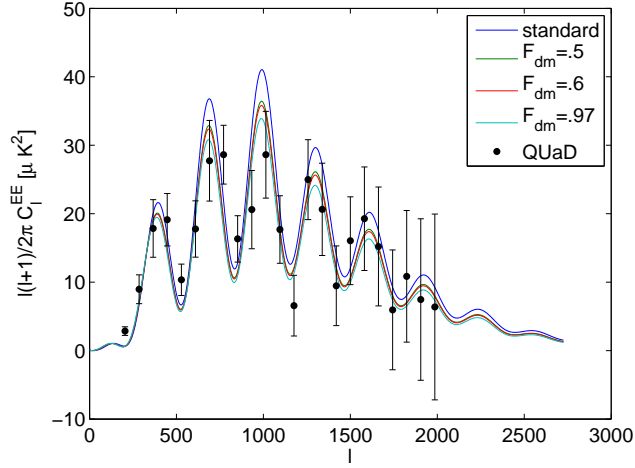}
\caption{Power spectrum of CMB E mode polarization}
\label{Cl_EE}
\end{figure}

\begin{figure}
\includegraphics[height=.3\textheight]{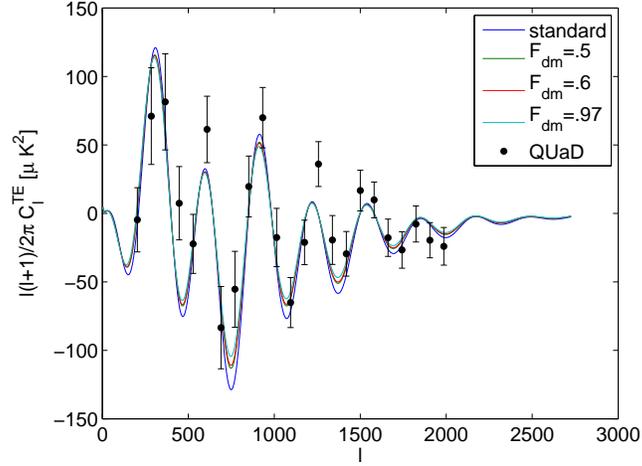}
\caption{Correlation of CMB temperature and E mode polarization}
\label{Cl_TE}
\end{figure}

As discussed in the previous sections, self-annihilation of dark matter affects ionization history, and therefore CMB anisotropy. 
By using \texttt{RECFAST} and \texttt{CAMB} code \cite{RECFAST1,RECFAST2,RECFAST3,CAMB} with small modifications, we have computed CMB power spectra and TE correlation, where self-annihilation of dark matter is considered. These are shown with the WMAP 5 year data \cite{WMAP5:basic_result,WMAP5:powerspectra}, ACBAR \cite{ACBAR,ACBAR2008} and QUaD data \cite{QUaD1,QUaD2,QUaD:instrument} in Fig. \ref{Cl_TT}, \ref{Cl_TE} and \ref{Cl_EE}.
As noticed in Fig. \ref{Cl_TT}, \ref{Cl_TE} and \ref{Cl_EE}, self-annihilation of dark matter particles decreases CMB anisotropy. 


\section{Analysis of the recent CMB data}
\label{analysis}
As discussed in the previous sections, self-annihilation of dark matter particles affects the CMB power spectrum and TE correlation.
Therefore, we may constrain self-annihilation (i.e. $F_{dm}$) with the recent CMB observations (WMAP + ACBAR + QUaD) \cite{WMAP5:basic_result,WMAP5:powerspectra,ACBAR,ACBAR2008,QUaD1,QUaD2,QUaD:instrument}).
For a cosmological model, we have considered $\Lambda$CDM + SZ effect + weak-lensing, where cosmological parameters are $\lambda_\alpha \in \left\{\Omega_b,\Omega_c,\tau,n_s, A_s, A_{sz}, H_0 , F_{dm} \right\}$. 
By using the \texttt{CAMB} and \texttt{CosmoMC} package \citep{CAMB,CosmoMC} with slight modifications, we have explored the likelihood $\mathcal{L}(C_l(\lambda_{\alpha})|C^{\mathrm{obs}}_l)$ in the multi-dimensional parameter space, 
where the CMB power spectra of given parameters, $C_l(\lambda_{\alpha})$, are compared with the observed power spectra  $C^{\mathrm{obs}}_l$. We have run the \texttt{CosmoMC} on a MPI cluster with 6 chains.
For the convergence criterion, we have adopted the Gelman and Rubin's ``variance of chain means'' and set the R-1 statistic to $0.03$ for stopping criterion \cite{Gelman:inference,Gelman:R1}. 

\begin{figure}[htb!]
\centering\includegraphics[scale=.5]{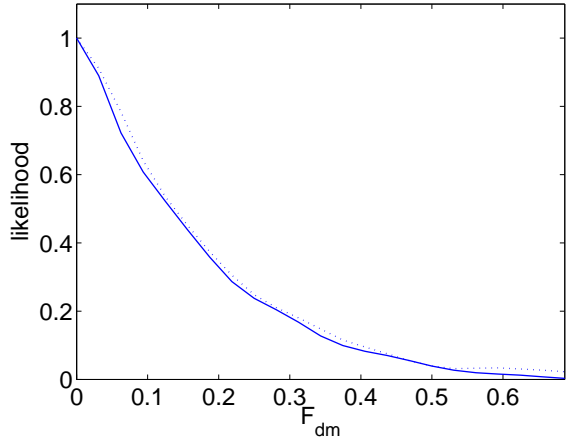}
\caption{Likelihood of $F_{\mathrm{dm}}$: normalized to its peak, a solid lines denote a marginalized likelihood and a dotted line a mean likelihood (refer to \cite{CosmoMC} for distinction between them).}
\label{quad1}
\end{figure}

\begin{figure}[htb!]
\includegraphics[scale=1]{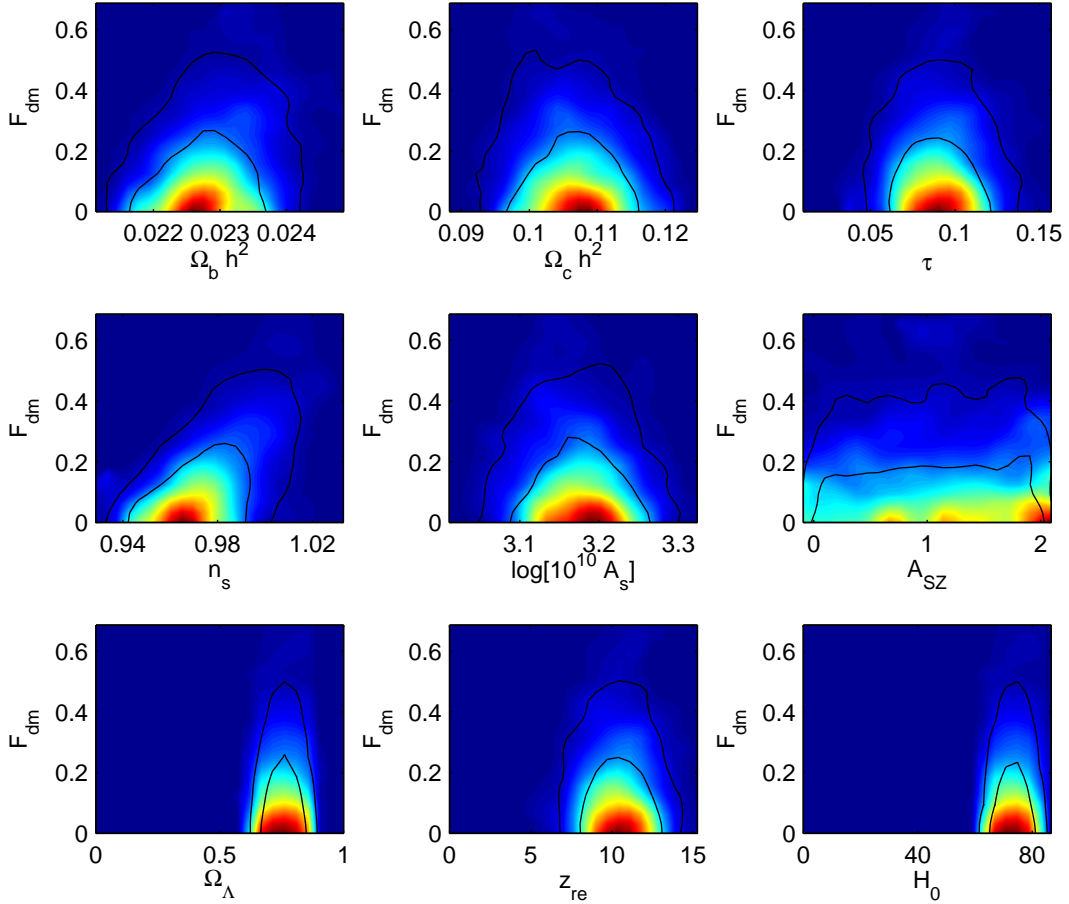}
\caption{The marginalized likelihood in the plane of $F_{\mathrm{dm}}$ versus others parameters. Two contour lines correspond to 1$\sigma$ and 2$\sigma$ levels respectively.}
\label{quad2}
\end{figure}

Analyzing the Monte-Carlo samples, we have obtained a posterior probability. From our posterior analysis, we impose an upper bound $F_{\mathrm{dm}}<0.7314$ at 95\% confidence level.
In Fig. \ref{quad1}, we show the likelihood distribution of $F_{\mathrm{dm}}$ parameter, marginalized over other parameters.
From Fig. \ref{quad1}, we may see that the CMB data favors $F_{dm}=0$.
In Fig. \ref{quad2}, we show the marginalized likelihoods in the plane of $F_{dm}$ versus other parameters.
Fig. \ref{quad2} also shows that $F_{\mathrm{dm}}=0$ is favored by the data.
As shown in Fig. \ref{quad1} and \ref{quad2}, we did not find statistically significant evidence on self-annihilation of dark matter particles.
In Table \ref{bestfit}, we summarize the best-fit values of cosmological parameters with 1 and 2 $\sigma$ confidence intervals respectively.

\begin{table}
\begin{tabular}{rrrrrrr}
\hline
symbol& description & best-fits &  lower (1$\sigma$) & upper (1$\sigma$)&  lower (2$\sigma$) & upper (2$\sigma$)\\
\hline
$\Omega_{b}\,h^2$ &  baryonic density $\times h^2$ &0.0227 & 0.0215 &  0.0243   & 0.021 & 0.0247\\
$\Omega_{c}\,h^2$  & cold dark matter density $\times h^2$ &0.1067 & 0.0926 & 0.1197 & 0.0898 & 0.1253\\
$\tau$ &optical depth    & 0.0963 & 0.0486 &  0.1355  & 0.043 & 0.1501\\
$F_{\mathrm{dm}}$  & annihilation of dark matter&0.0094 &  & 0.5933  &  & 0.7314\\
$n_s$ & spectral index & 0.9651 & 0.9356 & 1.0201  & 0.9313 & 1.0338\\
$\log[10^{10}A_s]$  & scalar amplitude  &3.1821 & 3.0595 & 3.2817 & 3.0094 & 3.3139\\
$A_{\mathrm{sz}}$  & SZ effect & 1.4166 & 0.0016 & 1.9945  &  & 1.9979\\
$z_{\mathrm{reion}}$ & re-ionization epoch & 11.0657 & 6.9958 & 13.6511  & 6.1755 &14.6326\\
$H_0$  [km/s/Mpc]& Hubble constant & 73.3856 & 67.3578 & 80.0397  & 65.7318 & 81.651\\
\hline
\end{tabular}
\caption{best-fit values of cosmological parameters}
\label{bestfit}
\end{table}

\section{Conclusion}
\label{conclusion}
By analyzing the recent CMB data, we have constrained the self-annihilation of dark matter particles.
We do not find statistically significant evidence on self-annihilation, and impose an upper bound on $F_{\mathrm{dm}}<0.7314$ at 95\% confidence level.
Due to the parameter degeneracy (i.e. $F_{dm}\propto\langle \sigma v \rangle/m_{\chi}$) in our analysis, significant self-annihilation 
is still possible, provided a dark matter particle is very massive (i.e. $m_{\chi}\gg 1\mathrm{GeV}$).
Therefore, a dark matter particle should be quite massive (i.e. $m_{\chi}\gg 1\mathrm{GeV}$), if the excess of energetic positrons and electrons in PAMELA/ATIC data is attributed to self-annihilation of dark matter particles.

Self-annihilation of dark matter particles leads to high level of gamma-ray emission from the region around the Galactic halo.
Therefore, Fermi Gamma-ray telescope will allow us to break some of parameter degeneracy and impose independent constraints on self-annihilation of dark matter.
Using the upcoming Planck data as well as Fermi Gamma-ray telescope data, we shall be able to impose important constraints on self-annihilation properties of dark matter particles.

\begin{theacknowledgments}
We acknowledge the use of the Legacy Archive for Microwave Background Data Analysis (LAMBDA), ACBAR, and QUaD data.
This work made use of the \texttt{CosmoMC} package and \texttt{RECFAST} code.
This work was supported by FNU grant 272-06-0417, 272-07-0528 and 21-04-0355. 

\end{theacknowledgments}



\bibliographystyle{aipproc}   

\bibliography{/home/tac/jkim/Documents/bibliography}

\end{document}